\documentclass[preprint,showpacs,preprintnumbers,amsmath,amssymb,pre]{revtex4}

\usepackage{graphicx}

\begin{document}

\preprint{preprint}

\title{A first--order irreversible thermodynamic approach to a simple energy converter}

\author{L. A. Arias--Hernandez}

\email{larias@esfm.ipn.mx}

\affiliation{Departamento de F\'{\i}sica, Escuela Superior de F\'{\i}sica y Matem\'aticas,
Instituto Polit\'ecnico Nacional, Edif. \# 9, U P Zacatenco, 07738, Ciudad de M\'exico, M\'EXICO. }

\author{R. T. Paez--Hernandez}

\email{phrt@correo.azc.uam.mx}

\affiliation{Area de F\'{\i}sica de Procesos Irreversibles, Depto. de CB, Universidad Aut\'{o}noma Metropolitana-A, Av. San Pablo \# 180, Ciudad de M\'exico, 02200, M\'{E}XICO}

\author{F. Angulo--Brown}

\email{angulo@esfm.ipn.mx}

\affiliation{Departamento de F\'{\i}sica, Escuela Superior de F\'{\i}sica y Matem\'aticas,
Instituto Polit\'ecnico Nacional, Edif. \# 9, U P Zacatenco, 07738, Ciudad de M\'exico, M\'EXICO. }

\date{\today}

\begin{abstract}
Several authors have shown that dissipative thermal cycle models based on Finite--Time Thermodynamics exhibit loop--shaped curves of power output versus efficiency, such as it occurs with actual dissipative thermal engines. Within the context of First--Order Irreversible Thermodynamics (FOIT),
in this work we show that for an energy converter consisting of two coupled fluxes it is also possible to find loop--shaped curves of both power output and the so--called ecological function against efficiency. In a previous work Stucki [J.W. Stucki, Eur. J. Biochem. \textbf{109}, 269 (1980)] used a FOIT--approach to describe the modes of thermodynamic performance of oxidative phosphorylation involved in $ATP$--synthesis within mithochondrias. In that work the author did not use the mentioned loop--shaped curves and he proposed that oxidative phosphorylation operates in a steady state simultaneously at minimum entropy production and maximum efficiency, by means of a conductance matching condition between extreme states of zero and infinite conductances respectively. In the present work we show that all Stucki's results about the oxidative phosphorylation energetics can be obtained without the so--called conductance matching condition. On the other hand, we also show that the minimum entropy production state implies both null power output and efficiency and therefore this state is not fulfilled by the oxidative phosphorylation performance. Our results suggest that actual efficiency values of oxidative phosphorylation performance are better described by a mode of operation consisting in the simultaneous maximization of the so--called ecological function and the efficiency.
\end{abstract}


\maketitle

\section{Introduction}
\label{intro}

It is well known that in actual dissipative heat engines, the experimental
plots of power output against thermal efficiency are loop--shaped
curves, where both the maximum power and maximum efficiency points
do not coincide and their separation can be managed by some phenomenological
parameters which depend on the engine's design and the materials employed
in its construction \cite{cgordon,gordon}. In these engines is common
to find irreversibilities (losses) due to friction and irreversible
heat fluxes. These losses must be taken into account in the models
elaborated to describe their general performance. In the thermal engine
modeling this kind of losses are usually considered in a separate
manner. However, one can couple the dissipative processes, in such
a way that loop--shaped curves are recovered. Se\-ve\-ral authors
\cite{gordon,anguloc} have proposed thermal engine models which reproduce
the loop--shaped curves observed in actual engines. This is accomplished
by means of characteristic functions (as power and efficiency) depending
on the thermal reservoir temperatures, and other parameters as compression
ratios, and thermal conductances for example. Through these quantities
one can control the distance between the maxima points of both power
output and efficiency.

In the present work (Section \ref{eff}), we show that in the case of a linear energy converter consisting of two coupled fluxes described by FOIT, in which one spontaneous flow manages a nonspontaneous one, loop--shaped curves (LSC) similar to those appearing in irreversible thermal engine models can be also obtained. These LSC for our FOIT--model are obtained for both power output ($P$) versus efficiency ($\eta$) and for ecological function ($E$) versus $\eta$, with ecological function defined as the power output minus the dissipation function ($T\sigma$, see
subsection \ref{chf}) \cite{angulo91}. In both loop-shaped curves it appears
a force ratio given by \cite{tribus}, 
\begin{equation}
x=\sqrt{\frac{L_{11}}{L_{22}}}\frac{X_{1}}{X_{2}},\label{x}
\end{equation}
(with $L_{ij}$ being the FOIT--phenomenological coefficients), which
measures a direct relationship between the two forces
$X_{1}$ and $X_{2}$ involved in the coupled flows. On the other hand, we also use
a coupling parameter $q$ given by,
\begin{equation}
q=\frac{L_{12}}{\sqrt{L_{11}L_{22}}}.\label{q}
\end{equation}
This parameter gives us a measure of the coupling of spontaneous and nonspontaneous fluxes \cite{caplan}. In section \ref{eff} we also show that by means of the LSC properties we can study several performance modes of the energy converter similar to those used in Finite Time Thermodynamics (FTT)\cite{calvo01, broeck05, jimenez06, jimenez07}, but with further results which consist in a set of functions describing how the maxima points can move one respect to the other. As an example of our previous results, we study the $ATP$ production occurring within mitochondria by means of respiration. A FOIT--approach to this problem was previously published by Stucki \cite{stucki80}. The Stucki's approach was mainly based on using two fluxes and two forces subject to several optimization criteria proposed by this author. The Stucki's results regarding the economic degrees of coupling of oxidative phosphorylation arise from the assumption that mitochondria is in a steady state corresponding to minimum entropy production which simultaneously corresponds to a maximum efficiency state. This last situation stems from the so--called conductance--mat\-ching condition (CMC), which is obtained by means of the inclusion of a third term in the expression for the entropy production, which corresponds to an attached cellular load under steady state conditions. The phenomenological coefficient of the third term mentioned can be fitted to obtain that the minima values of entropy production coincide with the optimum values of efficiency. Stucki uses several objective functions to model different mitochondria modes of operation but only by using the former two forces and two fluxes system and by assuming the implicit holding of the conductance--matching condition. In the present paper by means of the coupling of only two fluxes and two forces some of the main Stucki results are recovered without the inclusion of the third term in entropy production corresponding to the attached cellular load. This means that the attached load is not necessary to describe the energetics of a linear energy converter consisting in a pair of coupled processes where one drives the other. In summary, in this article we show that the minimum entropy production steady state regime is not equivalent to the optimum efficiency steady state regime. This assertion arises from the fact that the Stucki results are here obtained without the usage of the attached cellular load. If the Stucki's third term in the entropy production is considered, then the energetics formalism should be rewritten in terms of a 3 x 3 matrix of phenomenological coefficients, with its consequent changes in quantities such as the efficiency and the power output. This paper is organized as follows: In section \ref{eff} we discussed on the energetics of a \emph{2 x 2 system}; In section \ref{atp}, we applied some of the results of the previous section on the ATP--production under a linear approach. Finally in section \ref{con} we present some concluding remarks.

\section{On the energetics of a two--coupled fluxes system}
\label{eff}

In this section we study the following performance regimes: the minimum
dissipation function (MDF) \cite{prigogine}, the maximum power output
(MPO) \cite{odum55} and the maximum ecological function (MEF) \cite{angulo95,arias00}.
By means of these criteria we reproduce some previous results \cite{stucki80}
and we show some new ones. In particular, we will show that a linear
isothermal--isobaric engine working in the MEF--regime can reach an
efficiency until $\eta_{MEF}=0.75$.

\subsection{Constitutive equations}
\label{ceq}
\begin{figure}[t]
\begin{center}
\includegraphics[width=8.25 cm, height=8.25 cm]{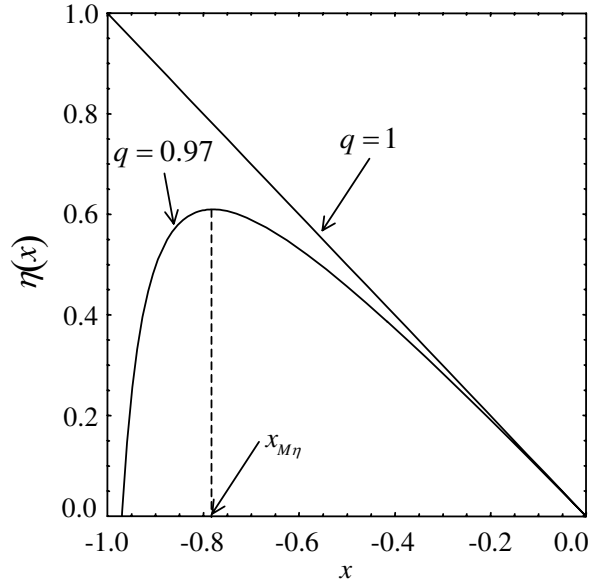}
\end{center}
\caption{Efficiency ($\eta$) vs. force ratio ($x$) for two different fixed coupling pa\-ra\-me\-ters. $x_{M\eta}$ corresponds to the maximum of $\eta$ for a fixed $q$.}
\label{figarias1}
\end{figure}
Let $J_{1}$ and $J_{2}$ be two coupled generalized fluxes (being
$J_{1}$ the driven flux and $J_{2}$ the driver flux); $X_{1}$ and
$X_{2}$ are the conjugate generalized potentials associated to the
fluxes. For the linear case, fluxes and potentials are after Onsager,
given by,
\begin{equation}
J_{1}=\sqrt{L_{11}}\left(\sqrt{L_{11}}X_{1}+q\sqrt{L_{22}}X_{2}\right)\label{J1q}\end{equation}
\begin{equation}
J_{2}=\sqrt{L_{22}}\left(q\sqrt{L_{11}}X_{1}+\sqrt{L_{22}}X_{2}\right),\label{J2q}\end{equation}
with $L_{12}=L_{21}$, the symmetry Onsager relation between crossed
coefficients. In these equations we use the coupling coefficient defined
by Eq. (\ref{q}) \cite{caplan}. Thus, in the limit case $q\rightarrow0$,
each flux is proportional to its proper conjugate potential through
its direct phenomenological coefficient, that is, the crossed effects
vanish, and therefore the fluxes become independent. When $q\rightarrow1$,
the fluxes tend to a mechanistic stoichiometry fixed relationship
independently of the potential magnitudes \cite{stucki80}.

On the other hand, it is convenient to define a parameter describing
the cross effect between both potentials. Taking $X_{2}>0$ as the
associated potential to the driver flux, we define the parameter
$x$ (see Eq. (\ref{x})) measuring the fraction of $X_{1}<0$ appearing due to the presence of the flux $J_{2}$. This
parameter is a quantity with values in the interval $\left[-1,0\right]$.
To complete the set of constitutive equations, we define the efficiency
of the thermodynamic process as follows,
\[
\eta=\frac{energy\: output}{energy\: input},
\]
and following Caplan and Essig \cite{caplan}, in terms of Onsager
relations we have,
\begin{equation}
\eta=-\frac{TJ_{1}X_{1}}{TJ_{2}X_{2}}=-\frac{\sqrt{L_{11}}\left(\sqrt{L_{11}}X_{1}+q\sqrt{L_{22}}X_{2}\right)X_{1}}{\sqrt{L_{22}}\left(q\sqrt{L_{11}}X_{1}+\sqrt{L_{22}}X_{2}\right)X_{2}}.\label{eftil}
\end{equation}
From Eq. (\ref{x}), we get an expression for $\eta$ in terms of
$x$ and $q$ as follows,
\begin{equation}
\eta\left(x,q\right)=-\frac{\left(x+q\right)x}{qx+1}.\label{efxq}
\end{equation}

A plot of $\eta$ versus $x$ for a fixed $q$ is depicted in Figure \ref{figarias1}. This graph is a convex curve with only a maximum point.
That is, there exists a relation between the input and output energetic
fluxes which maximize the efficiency for some given Onsager coefficients.
We take now a linear isothermal--isobaric conversion process and
build its characteristic functions in terms of $q$ and $x$ for determining
the performance conditions according to a certain objective function
maximization. First, we calculate the following characteristic functions: the dissipation function, the power output and the ecological function.

\subsection{Characteristic functions}
\label{chf}

\begin{figure}[t]
\begin{center}
\includegraphics[width=8.25 cm, height=8.25 cm]{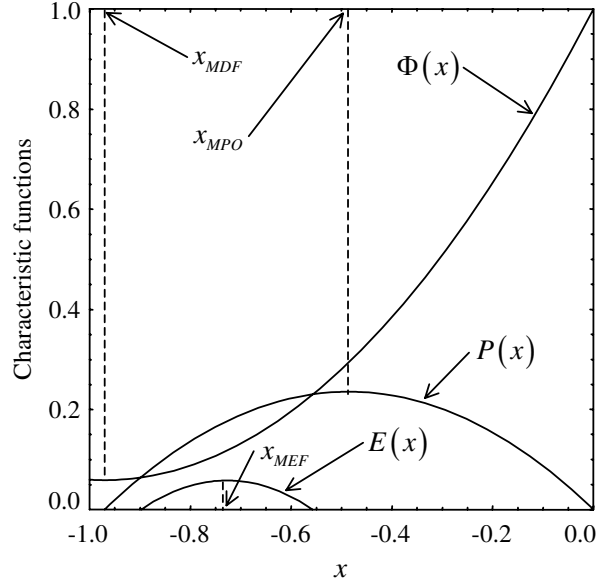}
\end{center}
\caption{Characteristic functions normalized respect to the fixed quantity $TL_{22}X_{2}^{2}$ versus the force ratio $x$. Dissipation function $\Phi\left(x\right)$, power output $P\left(x\right)$ and Ecological function $E\left(x\right)$, all of them for fixed $q=0.97$. $\Phi\left(x\right)$ reaches its minimum value at $x_{MDF}=-q$, $P\left(x\right)$ reaches its maximum value at $x_{MPO}=-\frac{q}{2}$ and $E\left(x\right)$ reaches its maximum value at $x_{MEF}=-\frac{3q}{4}$.}
\label{figarias2}
\end{figure}

\emph{Dissipation function}.- Within FOIT framework, for a \emph{2 x 2 system} of fluxes and forces the entropy production is given by \cite{caplan,degroot},
\begin{equation}
\sigma=J_{1}X_{1}+J_{2}X_{2},\label{dis}
\end{equation}
and following Tribus \cite{tribus}, the dissipation function for
an isothermal system can be expressed as $\Phi=T\sigma$. By the substitution
of Eqs. (\ref{J1q}) and (\ref{J2q}) into Eq. (\ref{dis}), in terms
of the parameters $x$ and $q$, $\Phi$ becomes,
\begin{equation}
\Phi\left(x,q\right)=\left(x^{2}+2xq+1\right)TL_{22}X_{2}^{2}.\label{disx}
\end{equation}
This function has a minimum value at the point $x_{MDF}=-q$ for
$q$ and $TL_{22}X_{2}^{2}$ fixed (see Figure \ref{figarias2}),
this means that the fraction of $X_{1}$ due to the presence of $X_{2}$
must be ${X_{1}}_{MDF}=-\left(\frac{L_{12}}{L_{11}}\right)X_{2}$, to obtain a
minimum dissipation steady state.

\emph{Power output}.- For isothermal processes
of two coupled fluxes the power output is given by \cite{odum55,angulo95}
\begin{equation}
P\left(x,q\right)=-TJ_{1}X_{1}=-x\left(x+q\right)TL_{22}X_{2}^{2}.\label{Pirr}
\end{equation}
This equation corresponds to a convex curve (Figure \ref{figarias2}),
with a maximum value at $x_{MPO}=-\frac{q}{2}$, this condition implies
that the fraction of the driven potential due to driver one in a
maximum power output regime must be ${X_{1}}_{MPS}=-\left(\frac{L_{12}}{2L_{11}}\right)X_{2}$.

\emph{The Ecological function}.- Defining the ecological function
as $E=P-\Phi$, by means of Eqs. (\ref{Pirr}) and (\ref{disx}),
we get
\begin{equation}
E\left(x,q\right)=-\left(2x^{2}+3xq+1\right)TL_{22}X_{2}^{2}.\label{Eirr}\end{equation}
This equation also corresponds to a convex curve with only a maximum
point (Figure \ref{figarias2}) at $x_{MEF}=-\frac{3q}{4}$, this
means that in the maximum ecological regime the relation between the
driver and driven flows must be ${X_{1}}_{MEF}=-\left(\frac{3L_{12}}{4L_{11}}\right)X_{2}$.
In this regime the conversion process undergoes a pathway accomplishing
a good compromise between power output and dissipated energy, that is, with a small decrement in $P$ we get a great decrement in $T\sigma$ \cite{angulo95,arias00}.

\subsection{Loop--shaped plots}
\label{loop}
\begin{figure}[t]
\begin{center}
\includegraphics[width=8.25 cm, height=8.25 cm]{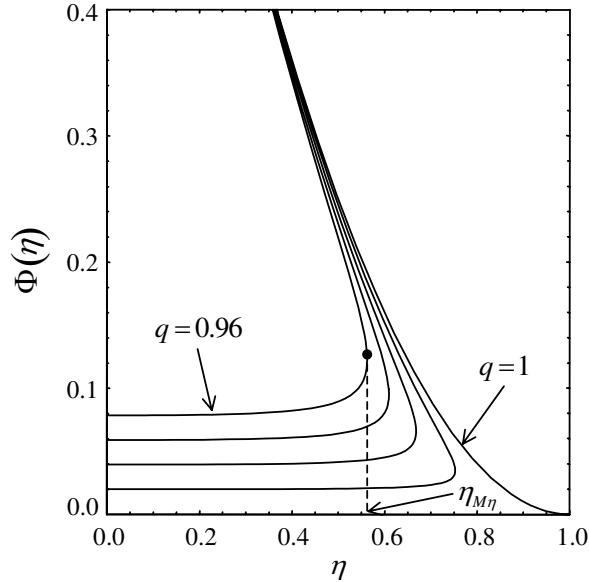}
\end{center}
\caption{Dissipation function ($\Phi$) versus efficiency ($\eta$) for $q=1,\:0.99,\:0.98,\:0.97,\:0.96$ respectively. $\eta_{M\eta}$ is the point which corresponds to the maximum efficiency for each case (here the case $q=0.96$ is marked).}
\label{figarias3}
\end{figure}

\begin{figure}[t]
\begin{center}
\includegraphics[width=8.25 cm, height=8.25 cm]{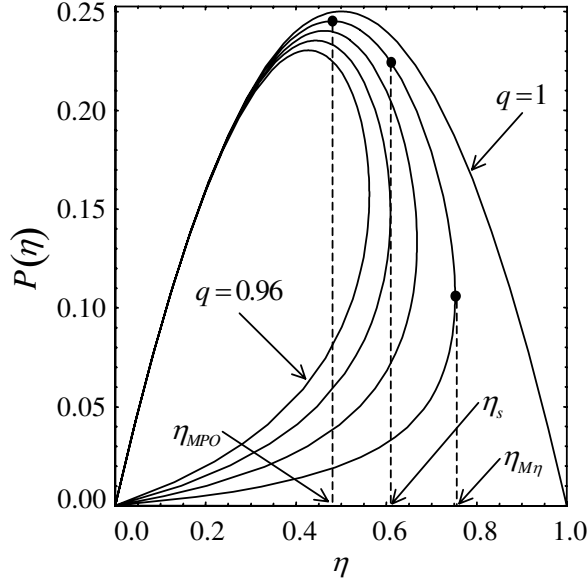}
\end{center}
\caption{Power output ($P$) versus efficiency ($\eta$) for the same $q's$ as in Figure 3. $\eta_{MPO}$ and $\eta_s$ are the points corresponding to the maximum power output and the semisum efficiency given by Eq. (\ref{ns}), respectively (here the case $q=0.99$ is marked).}
\label{figarias4}
\end{figure}

\begin{figure}[t]
\begin{center}
\includegraphics[width=8.25 cm, height=8.25 cm]{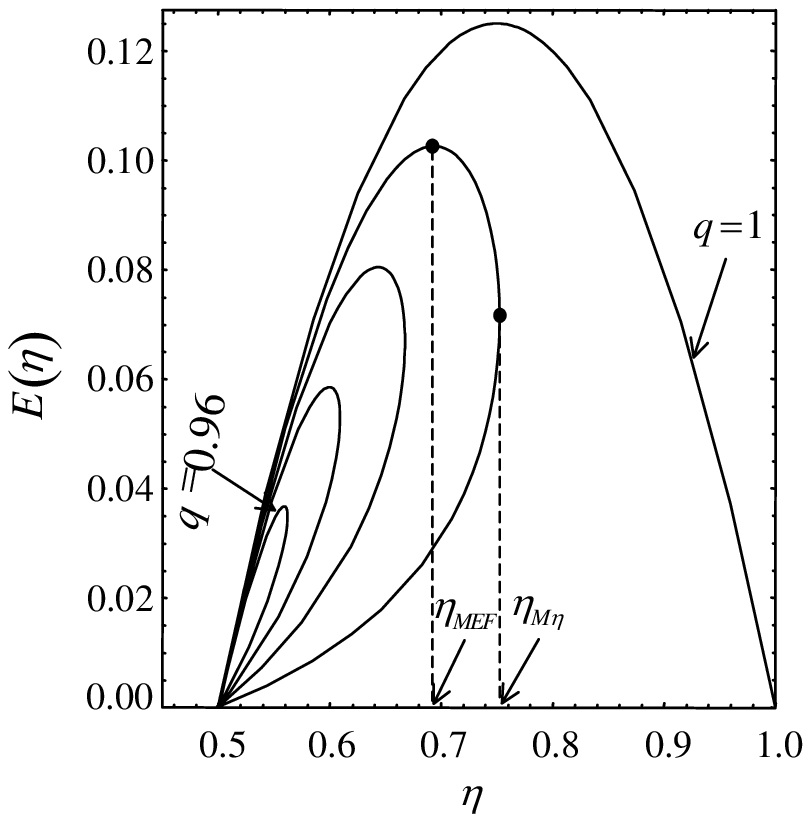}
\end{center}
\caption{Ecological function ($E$) versus efficiency ($\eta$) for the same $q's$ as in Figure 3. $\eta_{MEF}$ is the point which corresponds to the maximum ecological function (here the case $q=0.99$ is marked).}
\label{figarias5}
\end{figure}

Here, we find LSC by using FOIT-equations in an analogous way as it occurs in FTT--models and 
actual thermal engines. The functions $\eta$ (Eq. \ref{efxq}), $\Phi$ (Eq. \ref{disx}),
$P$ (Eq. \ref{Pirr}) and $E$ (Eq. \ref{Eirr}) depend on two parameters,
and three of them ($\eta$, $P$ and $E$) are convex functions with respect to $x$. From the plots corresponding to $\eta$ (Figure \ref{figarias1}) and $P$
(Figure \ref{figarias2}), we observe they have two zeros:
when $\sqrt{L_{11}}X_{1}=q\sqrt{L_{22}}X_{2}$, corresponding to a
first order steady state \cite{groot}, and when $\sqrt{L_{11}}X_{1}<<q\sqrt{L_{22}}X_{2}$
corresponding to a totally irreversible energy transfer, in both cases
$\eta=0$. We can transform $P$, $E$ and $\Phi$ as functions of
$\eta$ \cite{gordon,santillan97}. By means of Eq. (\ref{efxq}),
we first get $x\left(\eta,q\right)$ as
\begin{equation}
x\left(\eta,q\right)=-\frac{q\left(1+\eta\right)\pm{\sqrt{q^{2}\left(1+\eta\right)^{2}-4\eta}}}{2}.\label{xnq}
\end{equation}
Here, it is necessary to consider the two solutions of Eq. (\ref{xnq}),
because each one represents a branch of the plot $x$ versus $\eta$
for fixed $q$. By the substitution of Eq. (\ref{xnq}) into Eq. (\ref{disx}),
we get
\begin{equation}
\Phi\left(\eta,q\right)=\frac{\left(1-\eta\right)\left(2-q\left[q\left(1+\eta\right)\pm{R}\right]\right)}{2}TL_{22}X_{2}^{2},\label{dn}
\end{equation}
with $R=\sqrt{q^{2}\left(1+\eta\right)^{2}-4\eta}$. For the power output (Eq. \ref{Pirr}) we obtain
\begin{equation}
P\left(\eta,q\right)=\frac{\eta\left(2-q\left[q\left(1+\eta\right)\pm{R}\right]\right)}{2}TL_{22}X_{2}^{2},\label{pn}
\end{equation}
and in the same way, for the ecological function (Eq. \ref{Eirr}),
we have
\begin{equation}
E\left(\eta,q\right)=\frac{\left(2\eta-1\right)\left(2-q\left[q\left(1+\eta\right)\pm{R}\right]\right)}{2}TL_{22}X_{2}^{2}.\label{en}
\end{equation}
When we plot these functions against $\eta$ (Figures \ref{figarias3}, \ref{figarias4} and \ref{figarias5},
both branches) for $q\in\left[q_{min},1\right]$, we observe that $\Phi$ has a monotonically decreasing behavior, but $P$ and $E$ describe loop--shaped curves with some interesting points: the maxima $P$ and $E$ points ($\eta_{MPO}$ and $\eta_{MEF}$) and the maximum--$\eta$ points ($\eta_{M\eta}$, see Figures \ref{figarias4} and \ref{figarias5}). Plots in Fi\-gu\-res \ref{figarias4} and \ref{figarias5} are similar to those obtained in \cite{zhu} for a two--reservoir system with several irreversibilities. We will study these conspicuous points in next subsection.

\subsection{Performance modes and the coupling parameter}
\label{perf}

\subsubsection{Dissipation versus efficiency}
\label{dp}

In Figure \ref{figarias3}, we observe that while $q$ decreases
(i.e. the quality of the coupling diminishes), the minimum value of
$\Phi$ augments and $\eta_{M\eta}$ (see Eq. (16) below) also decreases.
This result may be the analogous of that occurring when the heat flux
through the body of a motor augments preventing that a spontaneous
flux can be used in managing a nonspontaneous one. In Figure \ref{figarias3},
we see that the minimum dissipation function occurs at $\eta_{MDF}=0$,
where $\Phi$ only depends on $q$ as
\begin{equation}
\Phi_{MDF}\left(\eta_{MDF}\right)=\left(1-q^{2}\right)TL_{22}X_{2}^{2},
\end{equation}
while the power output vanishes $P_{MDF}\left(\eta_{MDF}\right)=0$ \cite{santillan97nc}.
Evidently, when $q=1$, we recover the thermodynamic equilibrium state,
where all of the flows vanish, that is, to reach $\Phi_{MDF}\left(\eta_{MDF}\right)=0$
all interactions with the environment must be reversible processes.
For any other value of $q$, $\Phi_{MDF}\left(\eta_{MDF}\right)\neq0$.

\begin{figure}[t]
\begin{center}
\includegraphics [width=8.25 cm, height=8.25 cm]{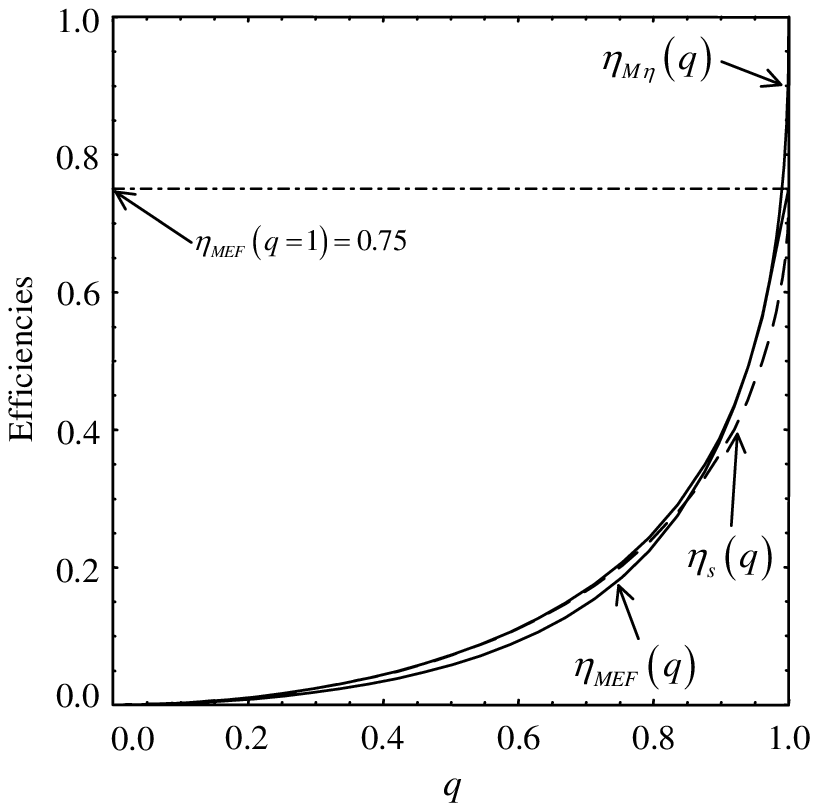}
\end{center}
\caption{Comparison between the efficiency of the M$\eta$--steady state (Eq.(\ref{ndmax})), the efficiency of the MEF--steady state (Eq.(\ref{nmef})) and the semisum efficiency $\eta_s$ (Ec.(\ref{ns}), dashed curve). $\eta_{MEF}\left( q=1\right) =0.75$ is the greatest efficiency of the MEF--regime.}
\label{figarias6}
\end{figure}

\begin{figure}[t]
\begin{center}
\includegraphics [width=8.25 cm, height=8.25 cm]{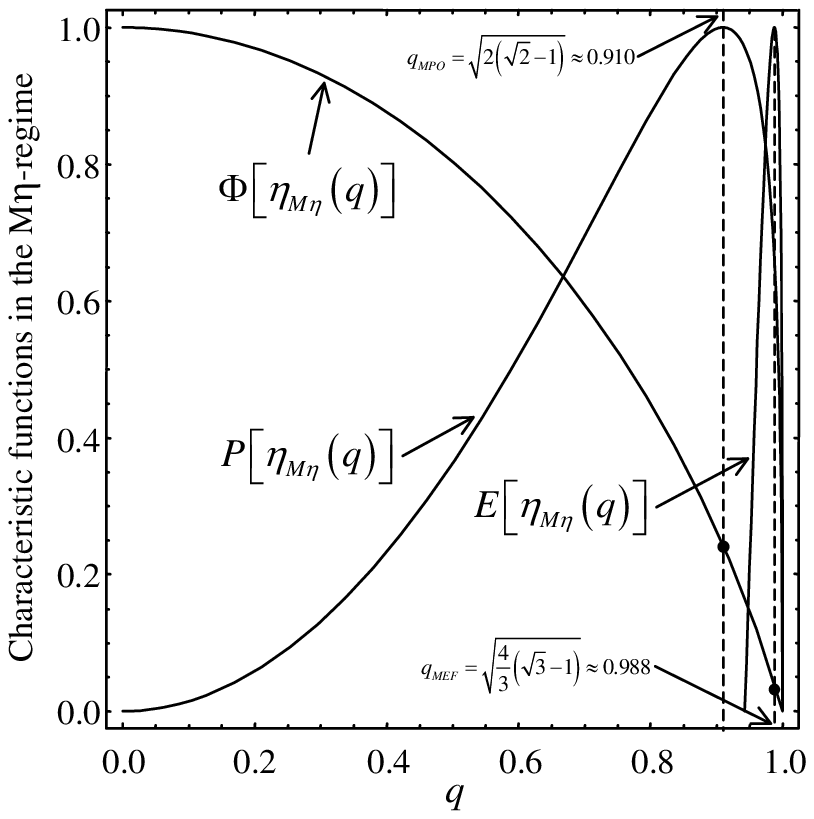}
\end{center}
\caption{Characteristic functions normalized respect to their maximum values in terms of the coupling parameter $q$, at the maximum efficiency regime. For the curve $\Phi\left[\eta_{M\eta}(q)\right]$ we can observe that $q_{MPO}\approx 0.910$ (where $P\left[\eta_{M\eta}(q)\right]$ attains its maximum value) does not correspond to a minimum entropy production steady state (with $q=1$), that is, the conductance--matching condition to reach the MPO--value at $\eta_{M\eta}$--regime is not necessary. Power output $P\left[\eta_{M\eta}(q)\right]$ and Ecological function $E\left[\eta_{M\eta}(q)\right]$ are also depicted. $q_{MEF}\approx 0.988$ corresponds to the maximum value of $E\left[\eta_{M\eta}(q)\right]$.}
\label{figarias7}
\end{figure}

Another point of interest is that where the efficiency reaches its
maximum value, which is found by means of $\left.\partial_{\Phi}\eta\right|_{\Phi_{\eta}}=0$,
and lead us to 
\begin{equation}
\eta_{M\eta}\left(q\right)=\frac{q^{2}}{\left(1+\sqrt{1-q^{2}}\right)^{2}},\label{ndmax}
\end{equation}
with a monotonically decreasing behavior while the irreversibilities
increase ($q\rightarrow0$, see Figure \ref{figarias6}
). By the substitution of Eq. (\ref{ndmax}) into Eqs. (\ref{dn}), (\ref{pn}) and (\ref{en}), we obtain only functions of $q$ such that $\Phi$ is monotonically decreasing with $q$, while $P$ and $E$ are convex functions with only a maximum point at $q_{MPO}=\sqrt{2\left(\sqrt{2}-1\right)}\approx 0.910$ and $q_{MEF}=\sqrt{\frac{4}{3}\left(\sqrt{3}-1\right)}\approx 0.988$,
respectively (see Figure \ref{figarias7}
). The first value was found by Stucki (see eq. 52 of \cite{stucki80}) by maximazing the objective function given by its equation 48 \cite{stucki80} (that is, the power output) subject to the CMC obtained from his assumption that the attached cellular load must be included in the entropy production (the mentioned third term, see eqs. 34, 35 and 36 of \cite{stucki80}). Nevertheless, here we show that this value ($q_{MPO}\approx 0.910$) can be obtained without considering any load, that is, by only optimizing the \emph{2 x 2 system} under maximum power output in the maximum efficiency--steady state. Then the minimum entropy production steady state is not equivalent to a maximum efficiency steady state (see Figure \ref{figarias7}). The second value ($q_{MEF}\approx 0.988$) has not an equivalent in the Stucki's treatment (see below and Figure\ref{figarias7}).

\subsubsection{Power output versus efficiency}
\label{pp}

The graph of $P\left(\eta,q\right)$ versus $\eta$ (see Figure \ref{figarias4})
shows that for $q=1$ a parabola is obtained. However, when we diminish
the quality of coupling, loop--shaped curves are obtained, in which
the relative position between the maxima points of power output and
efficiency respectively also depend on the coupling parameter $q$.
Besides, the parabola is the boundary mark of the loop--shaped curves,
that is, its maximum point gives the ideal power output when $q=1$,
which is $P\left[\eta_{MPO}\left(\left|q\right|=1\right)\right]=0.25\times TL_{22}X_{2}^{2}$.
This result coincides with that of ref. \cite{santillan97} for the
muscle contraction problem.

The efficiency which maximizes the power output is obtained by means
of $\left.\partial_{\eta} P\eta\right|_{\eta_{MPO}}=0$, giving
\begin{equation}
\eta_{MPO}=\frac{1}{2}\frac{q^{2}}{2-q^{2}}.
\label{nmpo}
\end{equation}
Here, we observe that only for $q=1$, $\eta_{MPO} =\frac{1}{2}$. At this point the process variables corresponding to the MPO--regime
satisfy the inequality $P_{MPO}=P\left[\eta_{MPO}\left(q\right)\right] \leq \Phi_{MPO}=\Phi\left[\eta_{MPO}\left(q\right)\right]$, that is,
\begin{equation}
\frac{q^{2}}{4}TL_{22}X_{2}^{2} \leq \left(1-\frac{3}{4}q^{2}\right)TL_{22}X_{2}^{2} \;\;\; \forall\; q\in\left[0,1\right],
\end{equation}
as it occurs in the MDF--regime with the advantage that in this MPO--regime the power output is not zero, as it is the case for the power output in the minimum entropy production steady state \cite{santillan97nc}.

The maximum efficiency point is found by using $\left.\partial_{P}\eta\right|_{P_{\eta}}=0$,
which also leads to Eq. (\ref{ndmax}), and therefore the behavior
of the process variables is the same as in the previous case.
In Figure \ref{figarias4}, we see that the bigger $q$ the smaller
$P\left(\eta_{M\eta}\right)$, that is, when $\eta_{M\eta}$ increases $P\left(\eta_{M\eta}\right)$ decreases until the limit case of the parabola ($\eta_{M\eta}=1$ and $P=0$). In Figure \ref{figarias4}, we also see
that the distance in the $\eta$--axis between the maximum power and the maximum efficiency points diminishes while the quality of the coupling diminishes and reaches a null value when $q=0$. In the following paragraph we will see how the ecological function gives a good tradeoff between maximum power and maximum efficiency.

\subsubsection{Ecological function versus efficiency}
\label{ep}

The MPO--regime provides a maximum energy output rate, but also produces
a great energy dissipation taken from the energy input (low efficiency).
Nevertheless, there exist some phenomena where this is not observed,
that is, the dissipation is always smaller than the power output.
In fact, many natural processes (biologic and nonbiologic) work following
a good compromise between $P$ and $\Phi$ \cite{caplan,calvo01,stucki80,angulo95,santillan97,lehninger,barranco,smith05,aledo04,lou02,pfieffer01,zhen99}.
On the other hand, as we see in Figure \ref{figarias5}, the ecological
function (Eq. (\ref{en})) plotted versus efficiency also gives
loop--shaped curves. Therefore, there exists an efficiency for which
one obtains the best compromise between $P$ and $\Phi$. This point
is found by means of $\left.\partial_{\eta} E\right|_{\eta_{MEF}}=0$,
which leads to 
\begin{equation}
\eta_{MEF}=\frac{3}{4}\frac{q^{2}}{4-3q^{2}}.
\label{nmef}
\end{equation}
This result gives a $q$ interval where the following inequality now
is satisfied $\Phi_{MEF}=\Phi\left[\eta_{MEF}\left(q\right)\right] \leq P_{MEF}=P\left[\eta_{MEF}\left(q\right)\right]$, that is,
\begin{equation}\hspace*{-0.35cm}
\left(1-\frac{15}{16}q^{2}\right)TL_{22}X_{2}^{2}\leq\frac{3}{16}q^{2}TL_{22}X_{2}^{2} \, \forall\, q\in\left(\frac{\sqrt{8}}{3},1\right].
\end{equation}
This inequality has some implications about the MEF--regime. In the
limit $q=1$, $\eta_{MEF}=0.75$, that is, $0.25$ more
than the MPO--efficiency, depending on the system's design. Other
result is that for any value of $q$, $P_{MEF}=0.75\times P_{MPO}$,
while the dissipation function in the case of MEF--regime
suffers a drastic decreasing compared with $\Phi_{MPO}$. In fact
we have
\begin{equation}
\Phi_{MEF}=\frac{1}{4}\left(\frac{16-15q^{2}}{4-3q^{2}}\right)\Phi_{MPO}.
\end{equation}
That is, within the $q$--interval where $E>0$ ($q\in\left(\frac{\sqrt{8}}{3},1\right]$),
$\Phi_{MEF}$ goes from $0.5\times\Phi_{MPO}$ to $0.25\times\Phi_{MPO}$.
To find the maximum efficiency we solve $\left.\partial_{E}\eta\right|_{E_{\eta}}=0$,
obtaining the same result that in the two previous regimes, that is,
the maximum efficiency has the same value for all of the performance
regimes given by Eq. (\ref{ndmax}). Therefore, we have only two $q$
values that maximize both the power output and the ecological function
in the M$\eta$--regime, that is, $q_{MPO}\approx 0.910$ and $q_{MEF}\approx 0.988$, respectively (see Figure \ref{figarias7}). This last value is bigger than the biggest $q$--value (at maximum efficiency) found by Stucki corresponding to the maximization of his function $J_1 X_1 \eta$  \cite{stucki80}. $q_{MEF}\approx 0.988$ was also found without using the CMC.

Additionally, in the loop--shaped curves of $P$ versus $\eta$,
we observe that it is possible to find an in\-ter\-me\-dia\-te point between
maximum power and maximum efficiency accomplishing a good compromise
between these two ways of performance \cite{angulo91}. This point is
\begin{eqnarray}
\eta_{s}\left(q\right)&=&\frac{1}{2}\left[\eta_{MAX}+\eta_{MPO}\right]\nonumber \\ &=&\frac{1}{2}\left[\frac{q^{2}}{\left(1+\sqrt{1-q^{2}}\right)^{2}}+\frac{1}{2}\frac{q^{2}}{2-q^{2}}\right].\label{ns}
\end{eqnarray}
If we compare this expression with the MEF--efficiency, $\eta_{MEF}$,
we obtain the behavior shown in Figure \ref{figarias6}, that is, $\eta_{MEF}\approx\eta_{s}$. This means that the $\eta$ values of the MEF--regime are a good compromise between power output and maximum efficiency, which is equivalent to a low dissipation regime, with the additional fact that this occurs for a realistic $q<1$ (for example, Stucki reported a $q_{exp}\approx 0.95$ for liver mitochondria from male rats \cite{stucki80}). However, some authors \cite{aledo04} have considered the ideal case $q=1$ to study some $ATP$ problems.

\section{On $ATP$ production: a linear approach}
\label{atp}

The purpose of this section is an application of the methodology presented
in Section \ref{eff}. One of the most important examples of the energy conversion in biology is the aerobic ATP--synthesis
(see references \cite{lehninger,smith05,aledo04,lou02,pfieffer01,zhen99}).
The global chemical reaction of $ATP$ synthesis is given by \cite{lehninger},
\begin{eqnarray}
\left\{ C_{6}H_{12}O_{6}+6\, O_{2}+6\, H_{2}O\right\} +\left[36\, ADP+36\, P^{+}\right] \nonumber \\ \rightleftharpoons\left\{ 6\, CO_{2}+12\, H_{2}O\right\} +\left[36\, ATP\right],
\label{r}
\end{eqnarray}
where driver and driven reactions have been indicated with curly and
square brackets, respectively. For the spontaneous reaction we take
$J_{2}X_{2}>0$ and for the nonspontaneous one, $J_{1}X_{1}<0$. Some
experiments suggest this process occurs out but near an equilibrium
state \cite{caplan,stucki80,lehninger}, thus, we can use the formalism
of Section \ref{eff}. Stucki considered as a reasonable idea that \emph{in vivo} oxidative
phosphorylation simultaneously operates at both maximum efficiency and minimum entropy production. This situation in Stucki's words is reached in a steady state named conductance matching
between extreme states of zero and infinite conductances respectively \cite{stucki80}. However, as it can be observed in Figure \ref{figarias4} (which stems from the parametric combination of Eqs. (\ref{efxq}) and (\ref{Pirr}) for realistic values of $|q|<1$), there exists a unique point with simultaneous zero values for $\eta$ and $P$ (which corresponds to the minimum entropy production steady state, see Figure \ref{figarias3}). Thus, for $q<1$ a conductance--matching condition in the Stucki's sense is not necessary. 
Among the four objective functions proposed in \cite{stucki80},
that given by $F_{S}=J_{1}X_{1}\eta$ at maximum efficiency, is in Stucki's words the
most suitable for the oxidative phosphorylation case. The optimal
efficiency obtained with this function is $\eta_{F}=0.618$ arising from a coupling parameter of $q_p^{ec}\approx 0.972$ (his Eq. (58)). This $\eta_{F}$ value is lower than reported efficiencies calculated from actual free energy changes, which are larger than their corresponding standard free energy changes for biochemical reactions as Eq.(\ref{r}). In fact, for $ATP$ synthesis under \emph{in vivo} conditions the efficiency can be around $0.736$ \cite{lehningerpb,torres}. Remarkably, these values $\eta_{F}=0.618$ and $q_p^{ec}\approx 0.972$ can be also obtained with the \emph{2 x 2} energy conversion formalim without using the CMC, in the following way: by the substitution of $\eta_{M\eta}$ given by Eq.(\ref{ndmax}) into the power output Eq. (\ref{pn}) and multiplying again by $\eta_{M\eta}$ we get $F_{S}\left(\eta_{M\eta}\right)$ (the so--called ``efficient power'' \cite{yilmaz1} in the maximum efficiency steady state), which corresponds to a convex curve with a maximum at $q\approx 0.972$ (i.e. $q_p^{ec}$). By substituting this $q$--value into Eq.(\ref{ndmax}) we obtain the same $\eta_{F}=0.618$ (without CMC). Thus, we obtained these results for $\eta_{M\eta}$ and $q_p^{ec}$, and that of subsubsection \ref{dp} for $q_p\approx 0.910$ by using only a \emph{2 x 2} formalism. In fact, all of the Stucki results of his Table I \cite{stucki80} for his four economic degrees of coupling can be obtained by the same procedure without recurring to the CMC.

If we take the efficiency value provided by the FOIT--formalism as a good criterion to choice the objective function at which oxidative phosphorylation thermodynamically performs,  then we can propose as objective function the ecological one due to its properties previously discussed. The efficient power ( $F_S = J_1 X_1 \eta$) used by Stucki \cite{stucki80} leads to $\eta_F=0.618$. However, Nelson and Cox (see Box 13.1 of \cite{lehningerpb}, pag. 498) reported that oxidative phosporylation under \emph{in vivo} conditions can reach efficiencies as high as $0.736$ due to actual free energy changes are larger than their corresponding standard free energy changes. If one observes Figure \ref{figarias4}, along the LSC between the maximum power output point and the maximum efficiency point, one has an infinite number of points corresponding each one of them to a particular mode of performance. Among these points one can choice some of them in terms of their energetic properties, such as a good compromise between high power output and low dissipation \cite{angulo91,arias97}. One mode of performance that accomplishes this goal is the so--called ecological function (see subsubsection \ref{ep}). If we use the ecological function  at maximum efficiency, which is analogous to the ``economic'' functions of Stucki, we find for the MEF--regime, $q_{MEF}\approx 0.988$ (see subsubsections \ref{dp} and \ref{ep}). If we substitute this value into Eq. (\ref{ndmax}) (the expression for the maximum efficiency) we get $\eta_{MEF}=0.732$, which is near estimated actual efficiency values \cite{lehningerpb,torres}. To compare the thermodynamic performance of the ecological function ($q_{MEF}\approx 0.988$) with the efficient power ($q^{ec}_p\approx 0.972$) we use a compromise function $C\left(q\right)$ of the type defined in \cite{arias97}. This function in terms of normalized quantities respect to the MPO--regime at optimum efficiency is given by
\begin{equation}
C\left(q\right)=\frac{P\left[\eta_{M\eta}\left(q\right)\right]}{P\left[\eta_{M\eta}\left(q_{MPO}\right)\right]}-\frac{\Phi\left[\eta_{M\eta}\left(q\right)\right]}{\Phi\left[\eta_{M\eta}\left(q_{MPO}\right)\right]}
\end{equation}
The function $C\left(q\right)$ has a maximum at $q_{C}\approx 0.982$, and it represents the best compromise between high power output and low dissipation. If this function is evaluated in $q_p^{ec}$, then $C(q_p^{ec})=0.479$. On the other hand, for $q_{MEF}$, $C\left(q_{MEF}\right)=0.489$, that is, a slightly larger value than $C\left(q_p^{ec}\right)$. However, in percentage terms $q_{MEF}$ is twice closer to $q_{C}\approx 0.982$ than $q_p^{ec}$. Thus, the ecological optimization provides a reasonable criterion for the thermodynamical performance of oxidative phosphorylation (without CMC), with the advantage of giving a high efficiency value within the range of actual values.

\section{Concluding remarks}
\label{con}

In the present paper we have developed a procedure to study an irreversible linear energy converter working under several steady--state conditions, by using optimization criteria stemming from some non--equilibrium approaches \cite{hoffmann}. Such criteria enhance the information given by the FOIT--formalim. In this way, one can describe the mutual influence between the operating mode of the energy converter (generalized forces and fluxes) and its design (phenomenological Onsager coefficients). Thus, we get some insights about the quantitative description of the energetics of a \emph{2 x 2 system}.

Our results permit to see that the FOIT--formalism along with FTT--procedures lead to loop--shaped curves such as it is observed in actual dissipative thermal engines \cite{gordon,anguloc} and some biological systems as it is the case of experimental data of efficiency versus power output reported by Smith et al \cite{smith05} for the soleous muscle of mouse. Our thermodynamic approach to the bridge between FOIT and some concepts arising from finite--time thermodynamics \cite{hoffmann} are somewhat different to that
suggested by Verhas and de Vos \cite{devos}.

Finally, in this work we have obtained all of the Stucki's results \cite{stucki80} for the optimal efficiency and the economic degrees of coupling of oxidative phosphorylation by using only a \emph{2 x 2} coupled system of fluxes without resorting the so--called conductance matching condition. Therefore, our approach suggests that the minimum entropy production regime is not compatible with the optimal efficiency steady state. In fact the MEP--regime leads to both zero efficiency and zero power output. On the other hand, we found that the so--called maximum ecological regime is suitable for the energetic description of oxidative phosphorylation, since this criterion gives a high efficiency value within the range of actual efficiencies. In addition, this regime represents a good compromise between high power output and low dissipation. In summary, our results indicate that the role played by the attached cellular load is not necessary for the energetic description of the two coupled fluxes involved in the biochemical reaction of $ATP$ synthesis, in the same way that no particular load is necessary to describe the internal energetics properties of a typical power plant. If one wishes to take into account the external load, a \emph{3 x 3} formalism is necessary for the overall thermodynamic description of the complete system, leading to a new set of energetic equations different to those as eqs. (\ref{eftil}) and (\ref{Pirr}). 

\begin{acknowledgments}
This work was supported by SIP, COFAA and EDI--IPN--M\'{E}XICO
and SNI--CONACyT--M\'{E}XICO.
\end{acknowledgments}

\end{document}